\documentclass[aps,prb,superscriptaddress,citeautoscript,twocolumn,notitlepage,longbibliography]{revtex4-2}

\usepackage{graphicx}
\usepackage{dcolumn}
\usepackage{bm}
\usepackage{svg}

\usepackage{graphicx}
\usepackage{amsmath}
\usepackage{amssymb}
\usepackage[normalem]{ulem}
\usepackage{braket}
\usepackage{color}
\usepackage{gensymb}
\usepackage[colorlinks=True,linkcolor=red,citecolor=blue,urlcolor=blue]{hyperref}

\usepackage{makecell}
\usepackage{dsfont}
\usepackage{mathtools,esint}
\newcommand{\bea}{\begin{eqnarray*}}
\newcommand{\eea}{\end{eqnarray*}}
\newcommand{\bne}{\begin{equation*}}
\newcommand{\ede}{\end{equation*}}

\newcommand{\bnen}{\begin{equation}}
\newcommand{\eden}{\end{equation}}
\newcommand{\bean}{\begin{eqnarray}}
\newcommand{\eean}{\end{eqnarray}}
\newcommand{\bsen}{\begin{subequations}}
\newcommand{\esen}{\end{subequations}}

\newcommand{\bna}{\begin{array}}
\newcommand{\eda}{\end{array}}
\newcommand{\bnm}{\begin{enumerate}}
\newcommand{\edm}{\end{enumerate}}
\newcommand{\bni}{\begin{itemize}}
\newcommand{\edi}{\end{itemize}}

\renewcommand{\vec}[1]{\text{\boldmath{$ #1 $}}}

\DeclareMathAlphabet\mathbfcal{OMS}{cmsy}{b}{n}

\newcommand{\orcidicon}[1]{\href{https://orcid.org/#1}{\includegraphics[height=\fontcharht\font`\B]{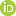}}}

\begin{document}

\title{Weyl-point teleportation}

\author{Gy\"orgy Frank\,\orcidicon{0000-0003-2129-2105}}

\affiliation{Department of Theoretical Physics,
Budapest University of Technology and Economics, 
Hungary}

\affiliation{MTA-BME Exotic Quantum Phases 
 Group,
 Budapest University of Technology and Economics, Hungary}

\author{D\'aniel Varjas\,\orcidicon{0000-0002-3283-6182}}
\affiliation{Department of Physics, Stockholm University, AlbaNova University Center, 106 91 Stockholm, Sweden}

\author{Gerg\H{o} Pint\'er\,\orcidicon{0000-0003-1289-0592}}

\affiliation{Department of Theoretical Physics,
Budapest University of Technology and Economics, 
Hungary}

\affiliation{MTA-BME Exotic Quantum Phases 
 Group,
 Budapest University of Technology and Economics, Hungary}

\author{Andr\'as P\'alyi}

\affiliation{Department of Theoretical Physics,
Budapest University of Technology and Economics, 
Hungary}

\affiliation{MTA-BME Exotic Quantum Phases 
 Group,
 Budapest University of Technology and Economics, Hungary}

\date{\today}

\begin{abstract}
In this work, we describe the phenomenon of Weyl-point 
teleportation. Weyl points usually move continuously
in the configuration parameter space of a quantum system
when the control parameters are varied continuously. 
However, there are special 
transition points in the control space 
where the continuous motion of the Weyl points is disrupted.
In such transition points, an extended nodal structure
(nodal line or nodal surface) emerges, serving as a 
wormhole for the Weyl points, allowing their teleportation
in the configuration space.
A characteristic side effect of the teleportation is that the motional susceptibility
of the Weyl point diverges in the vicinity of the transition point, 
and this divergence is characterized by a universal scaling
law. We exemplify these effects via a two-spin model and a Weyl Josephson circuit model.
We expect that these effects generalize to many other settings including electronic band structures of topological semimetals. 
\end{abstract}

\maketitle

\emph{Introduction.}
Electronic band structures of crystalline materials 
often exhibit band touching points \cite{Neumann,Herring,Armitage}, leading to 
characteristic phenomena such 
as surface Fermi arcs~\cite{Huang2015}, chiral anomaly, anomalous Hall effect, quantum oscillations, and quantized circular photogalvanic effect~\cite{dejuan2017, Ma2017}. 
The generic pattern for band touching
is the Weyl point, where the dispersion relation in the vicinity
of the degeneracy point starts linearly in all directions. 

However, for many materials, the spatial symmetry of the crystal 
structure implies that band touching can 
happen in the form of extended nodal structures, 
such as nodal lines or nodal surfaces \cite{BeriPRB2010,CarterPRB2012,ChenFang,WeikangWu,ChenFangChinese2016,bzduvsek2016nodal,QiFengLiangPRB2016,YingMingXie,Zhao,Zhang,Chan,Hirschmann,Wilde,GonzalezHernandez, Yang2017}. Furthermore, nodal lines and nodal surfaces
can also arise in 
parameter-dependent quantum systems \cite{Frank,FrankDensity,Fatemi} in the presence of fine-tuning or symmetries.
For brevity, we will use `fine-tuned' to describe both
of these scenarios. 

Generic perturbations with respect to the fine-tuned 
setting will necessarily dissolve the extended
nodal structures into a number of Weyl points.
For example, mechanical strain can break the symmetry 
of a crystal and thereby split a nodal line into Weyl points \cite{YingMingXie}.
Details of such a dissolution process have various physical implications, e.g., a qualitative change in the density of states and the properties of the surface states, etc. In this work we uncover general properties of this dissolution process.  

In the above setting, the six independent components of the mechanical strain tensor provide examples of \emph{control parameters}.
We will refer to the fine-tuned control parameter point where the extended nodal structure appears as the \emph{transition point}.
Furthermore, we will call the parameter space in which 
the degeneracy structures live, i.e., the
three-dimensional momentum space in the above example, 
the \emph{configuration space}.

In this work, our main observation is that a nodal loop or nodal surface emerging at a transition
point can be thought of as a `wormhole' for Weyl points, allowing for the `teleportation' of Weyl points.
A side effect of teleportation is that the motion of the Weyl points become singular as the control parameters approach the transition point: an infinitesimal change in the control parameters
induces a macroscopic displacement of the Weyl point in the configuration space. 

\begin{figure*}
	\begin{center}
		\includegraphics[width=2.0\columnwidth]{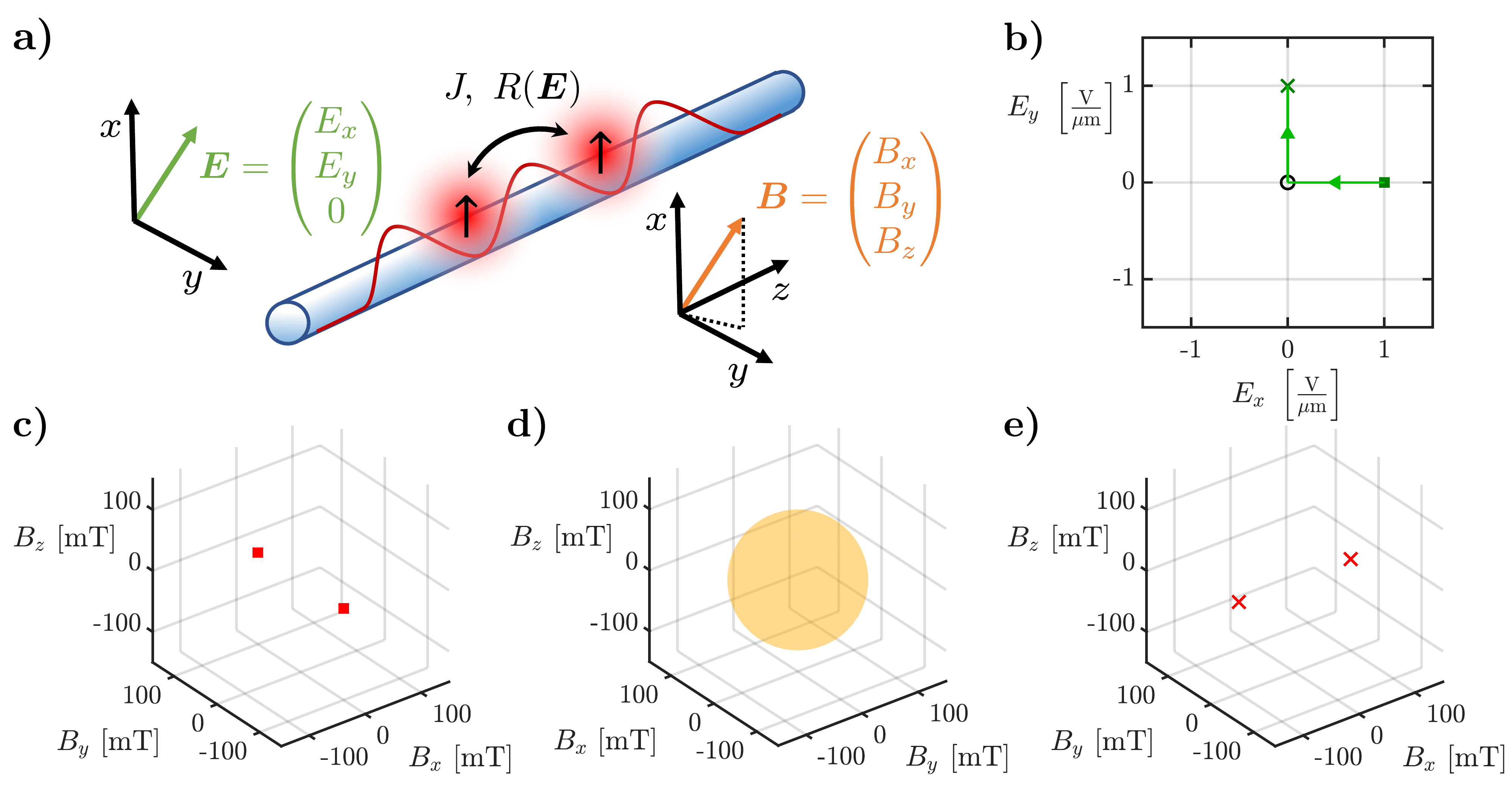}
	\end{center}
	\caption{{\bf Teleportation of magnetic Weyl points of a spin-orbit-coupled
double quantum dot.} 
	(a) Double quantum dot hosting two interacting electrons, 
	in the presence of a Zeeman field $\vec B$, and spin-orbit interaction induced by the electric field $\vec E$. 
	(b) Trajectory of the control parameter vector, i.e., the electric field in the xy plane. 
	The trajectory is parametrized by $t \in [-1,1]$ with $\epsilon=1\text{V}/\mu\text{m}$ (see text).
	(c,d,e) Evolution of ground-state Weyl points in the 
	configuration space ($\vec B$ space), 
	as the control parameter vector
	is varied according to panel (b). 
	Parameters: $g=2$, $J=11.6 \, \mu\text{eV}$.
	(c) Weyl points for $t<0$.
	(d) Nodal surface (sphere) at $t=0$ (black circle in panel b).
	Teleportation of Weyl points happens at this point
	of the control-parameter trajectory. 
	(e) Weyl points for $t>0$.
	\label{fig:doubledot}}
\end{figure*}

\emph{Teleportation of magnetic Weyl points of a spin-orbit-coupled
double quantum dot.}
We illustrate Weyl-point teleportation using an elementary model of a two-electron double quantum dot with spin-orbit interaction. 
The setup is shown in Fig.~\ref{fig:doubledot}a.
It consists of a cylindrically symmetric semiconducting nanowire (blue) where
gate electrodes create a double-well potential (solid red),
with each well capturing a single spinful electron (red clouds). 
The two localized spins interact with each other via
exchange interaction. 

We assume that an external homogeneous electric field 
$\vec E = (E_x, E_y, 0)$ 
breaks the cylindrical symmetry of the
system, and induces an anisotropy in the Heisenberg interaction among the spins via the spin-orbit interaction of the Rashba type.
The two-dimensional space of $\vec E$ serves as the
control space.
A homogeneous external magnetic field $\vec B$ 
is also present in the setup,
taken into account by a Zeeman interaction term in the model. 
The three-dimensional space of $\vec B$ serves as the
configuration space.

The two-electron system is described by the following $4\times 4$ 
Hamiltonian:
\begin{equation}\label{eq:4x4}
    \hat{H}_{\text{DD}} = g \mu_B \vec B \cdot (\hat{\vec S}_L + \hat{\vec S}_R)
    + J \hat{\vec S}_L \cdot R(\vec E) \hat{\vec S}_R.
\end{equation}
Here $\hat{\vec S}_{L/R}$ is the spin operator of the left/right electron,
$g$ is the $g$-factor of the electrons,
$\mu_B$ is the Bohr magneton, 
$\vec B$ is the external magnetic field, 
$J$ is the exchange interaction strength, 
and $R(\vec E)$ is a $3\times 3$ rotation matrix describing the
effect of spin-orbit interaction on the exchange interaction \cite{Kavokin,Scherubl}.

The exchange term in Eq.~\eqref{eq:4x4} can be derived from the Rashba spin-orbit Hamiltonian and the corresponding two-site Hubbard model. 
Here, we outline the simple physical picture that predicts the qualitative dependence of the rotation matrix $R$ on the electric field $\vec E$. Assume that the electric field creates a spin-orbit term
of the Rashba type, $H_\text{so} = \alpha\hat{\vec S} \cdot \left(
\vec E \times \hat{\vec p}\right)$ \cite{ScherublRashba}.
When an electron tunnels from the right dot to the left ($\vec p || -\vec e_z$), it feels a spin-orbit magnetic field $\vec B_{\text{so}}\propto \vec e_z\times\vec E$, and hence its spin rotates around $\vec B_\text{so}$ with an angle $\theta=\gamma E$ proportional to the electric field causing the spin-orbit coupling. 
This effect is incorporated in Eq.~\eqref{eq:4x4} as the matrix $R$:
\bean
\label{eq:exchangerotation}
    R(\vec E) &=&
    \exp{
    \left[\gamma
    \begin{pmatrix}
    0&0&E_x\\
    0&0&E_y\\
    -E_x&-E_y&0
    \end{pmatrix}
    \right]
    }.
\eean

Without spin-orbit interaction (that is, $\vec E=0$), the system is isotropic. At zero magnetic field the ground state is a singlet, and the excited states are three degenerate triplets. Switching on an external magnetic field leaves the energy of the singlet state unchanged, but lowers the energy of one triplet state. At $B_0=J/\mu_{g \text B}$ the ground state becomes degenerate. In the magnetic parameter space, the ground state degeneracy points form a nodal surface, i.e., a spherical surface with radius $B_0$, see Fig.~\ref{fig:doubledot}d.

With spin-orbit interaction ($\vec E \neq 0$), the ground state degeneracies split everywhere except the two points where the external magnetic field $\vec B$ is parallel to the spin-orbit magnetic field $\vec B_\text{so}$.
These two remaining degeneracy points are located at 
\bean
\label{eq:doubledotweylpoint}
\vec B_{W,\pm}(\vec E)=\pm B_0\frac{\vec e_z\times\vec E}{E},
\eean
where $\vec{e}_z$ is the z-directional unit vector.

The above observations naturally combine into the
Weyl-point teleportation effect, as illustrated by
Fig.~\ref{fig:doubledot}b-e.
Consider the continuous trajectory of the control vector shown in 
Fig.~\ref{fig:doubledot}b, parametrized by 
$t \in [-1,1]$.
This trajectory is defined as
$\vec E(t) = -t \, \epsilon \, (1,0,0)$ for $t<0$ and $\vec E(t) = t \, \epsilon \, (0,1,0)$ for $t \geq 0$, with $\epsilon > 0$ having the dimension of electric field. This trajectory contains the transition point $\vec E(0) = \vec 0$.
For $t<0$, Weyl points are located in $\pm B_0 \vec{e}_y$ (panel b), whereas for $t>0$, Weyl points are located in $\pm B_0 \vec{e}_x$ (panel e), 
and for $t=0$ there are no Weyl points but a 
degeneracy sphere (panel d) that serves as a `wormhole', allowing the 
`teleportation' of the Weyl points as the control-space
trajectory traverses the transition point. 

Note that in our model, the Weyl points do not move
`before' ($t<0$) and `after' ($t>0$) the teleportation.
This changes in more realistic models, e.g., taking into account the electric-field dependence of the $g$-factor; see also our discussion on Weyl Josephson circuits below.

\emph{Divergence of motional susceptibility.}
Here, we exemplify that the motion of Weyl points
in the configuration space 
is singular 
as the control parameters approach the transition point.
We use the same two-spin model as above. 

First, we introduce the \emph{motional susceptibility matrix} $\chi$ of a Weyl point, which is the quantity characterizing the motion of the Weyl point in configuration space in response to a small change of the control parameters. This matrix depends on the control parameters, and 
it is defined everywhere in the control space
except in the transition point.
For a selected Weyl point having the location $\vec B_\text{W}(\vec E)$ in the configuration space (see, e.g., Eq.~\eqref{eq:doubledotweylpoint}), it is defined as 
\begin{equation}
\label{eq:susceptibility}
    \chi_{ik}(\vec E)
    = 
    \frac{\partial B_{\text{W},i}(\vec E)}{\partial E_k},
\end{equation}
where $i \in \{x,y,z\}$ and $k \in \{x,y\}$.

Hence for our two-spin model, this susceptibility matrix has
dimension $3 \times 2$. 
It can be 
obtained analytically from Eq.~\eqref{eq:doubledotweylpoint}, e.g, by selecting the Weyl point $\vec{B}_{W,+}$, 
via the definition \eqref{eq:susceptibility} as
\begin{equation}
    \chi(\vec E) = \frac{B_0}{E^3}\begin{pmatrix}
    -E_x E_y & E_x^2\\
    -E_y^2 & E_x E_y\\
    0 & 0
    \end{pmatrix}.
\end{equation}
The singular value decomposition of this matrix
has the form 
$\chi = U \Sigma V^{\text T}$, 
where 
\begin{eqnarray}
U(\vec E) &=& \frac{1}{E}
\begin{pmatrix}
-E_x & E_y & 0\\
-E_y & -E_x & 0\\
0 & 0 & E
\end{pmatrix}\\
\Sigma (\vec E)&=&
\left(\begin{array}{cc}
     \Sigma_1 & 0 \\
     0 & \Sigma_2 \\
     0 & 0
\end{array}\right) 
\\
V (\vec E)&=& \frac{1}{E}
\begin{pmatrix}
-E_y&-E_x\\
E_x&-E_y
\end{pmatrix}
\end{eqnarray}
Here, the two singular values are $\Sigma_1 = B_0/E$, 
$\Sigma_2 = 0$.

Remarkably, the largest singular value $\Sigma_1$
diverges as $1/E$ as the control parameters approach
the transition point $\vec{E}=0$.
This shows that for paths that go close, but not through the transition point,
an infinitesimal change of the control parameter yields a large movement of the Weyl point in the configuration space.
This becomes a macroscopic jump when the path passes through the transition point and teleportation happens.

\begin{figure*}
	\begin{center}
		\includegraphics[width=2.0\columnwidth]{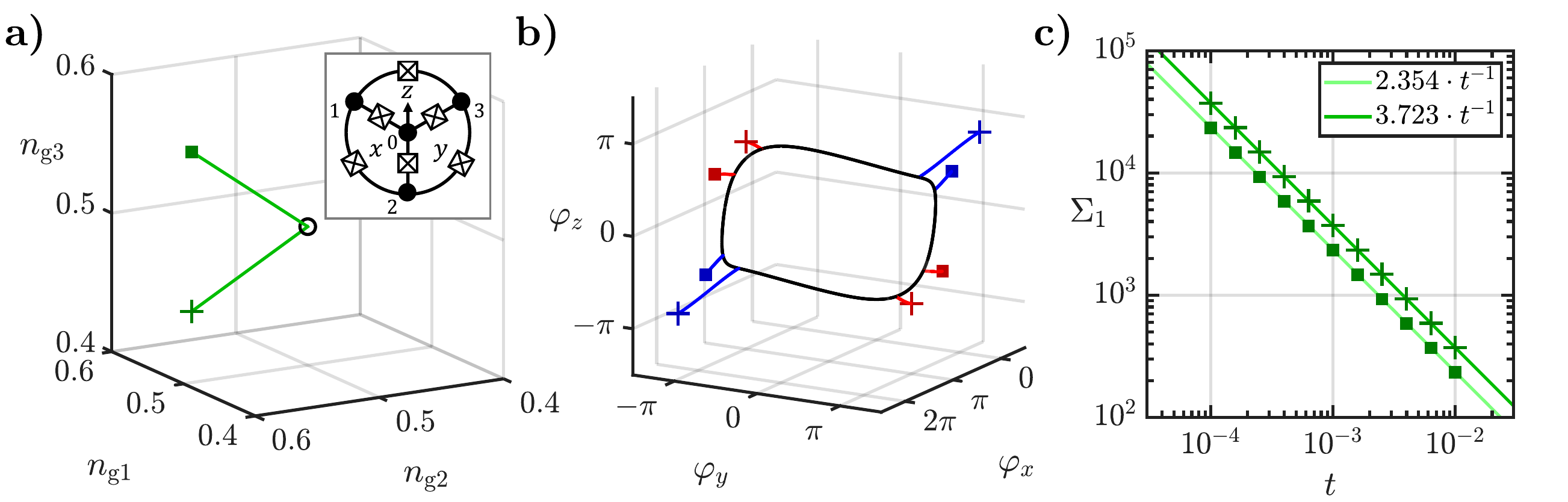}
	\end{center}
	\caption{{\bf Weyl points and their teleportation in a Weyl Josephson circuit.}
	(a) Continuous trajectory 
	$\blacksquare \to \circ \to +$
	of the control parameter vector, i.e., the vector $\vec{n}_\text{g}$ of gate voltages that controling the Weyl Josephson circuit proposed in \cite{Fatemi} (inset).
	(b) Evolution of four Weyl points in the configuration space, i.e., the space of the flux vector $\vec{\varphi}$, as the control parameter is varied on the path 
	$\blacksquare \to \circ \to +$
	in (a).
	Teleportation of the Weyl points happens at
	$\vec n_\text{g} = (0.5,0.5,0.5)$, marked by the black circle in (a), when the degeneracy points in the configuration space form a nodal loop (black) in panel (b).
    See text for parameters.
    (c) Scaling of Weyl-point motion in a Weyl Josephson circuit.
     The greatest singular value $\Sigma_1$
    of the motional susceptibility matrix
    $\chi$ exhibits a $1/t$ divergence
    as the control vector approaches the transition point as $\vec{n}_\text{g}(t) = \vec n_{\text{g}0} + t \vec{e}$.
    The two data sets correspond to 
    $\vec e = \vec e_{\blacksquare}$
    and
    $\vec e = \vec e_{+}$ (see text), and
    the motion of the top left Weyl point in panel b).
    The straight lines in c) are single-parameter fits to the data, see inset. 
	\label{fig:mercedes}}
\end{figure*}

\emph{Weyl Josephson circuits.}
To show the generic nature of the effects
discussed above, we now identify them in a different
setup: a Weyl Josephson circuit \cite{Fatemi}. 
The inset of Fig.~\ref{fig:mercedes}a shows 
the schematic arrangement of a 
multi-terminal Josephson circuit, originally proposed in 
Fig.~1 of \cite{Fatemi}.
It is built from four superconducting terminals
(black circles),
where terminal 0 is grounded, and terminals 1, 2, 
and 3 are floating and controlled by local gate
electrodes. 
The corresponding dimensionless gate voltages are denoted by
$n_{\text{g}\alpha}$ with $\alpha \in \{1,2,3\}$. 
We regard these gate voltages as the control 
parameters. 
Furthermore, the three loops 
formed by the terminals,
denoted as $x$, $y$, $z$ in Fig.~\ref{fig:mercedes}a,
are 
pierced by controllable magnetic fluxes
$\varphi_x$, $\varphi_y$, $\varphi_z$; 
we consider the 3D space of these fluxes as the
configuration space. 
We denote the Josephson energy (capacitance)
associated to the junction between terminals
$\alpha$ and $\beta$ as $E_{\text{J}, \alpha \beta}$ 
($C_{\alpha \beta}$).

The Hamiltonian of this Josephson circuit 
reads
\bean
\hat{H} &=&
E_{\text C}
\left(\hat{\vec n}-\vec n_{\text g}\right)
\cdot 
c^{-1}
\left(\hat{\vec n}-\vec n_{\text g}\right)\nonumber\\
&-& 
\sum\limits_{\substack{\alpha,\beta=0 \\ \alpha<\beta}}^{3} 
E_{\text{J}, \alpha \beta} 
\cos\left[\hat{\varphi}_\alpha-\hat{\varphi}_\beta+\gamma_{\alpha \beta}(\varphi_x, \varphi_y,\varphi_z)\right].
\eean
Here, $\hat{n}_\alpha$ is the number operator counting the Cooper pairs on terminal $\alpha \in \{0,1,2,3\}$,
and $\hat{\vec n} = (\hat{n}_1,\hat{n}_2,\hat{n}_3)$.
Furthermore, 
$\hat{\varphi}_\alpha$ are the phase operators canonically conjugated to $\hat{n}_\alpha$,
$E_{\text C}=(2e)^2/(2C_0)$ is a capacitance scale characteristic of the network of terminals, $c=C/C_0$ is the dimensionless capacitance matrix defined from the capacitance matrix \cite{vanderWielRMP} $C$ (see Appendix), and $\gamma_{\alpha \beta}$ are control angles depending on the fluxes as
$\gamma_{0\beta}=0$,
$\gamma_{12}=\varphi_x$,
$\gamma_{13}=-\varphi_z$,
and $\gamma_{23}=\varphi_y$.

If all three gate voltages $n_{\text{g}\alpha}$ have
integer or half-integer values, then the Hamiltonian
$\hat{H}$ above has an effective time-reversal symmetry (see Appendix),
implying that the ground-state degeneracy points,
if they exist, form a loop in the 
configuration space.
This is exemplified in Fig.~\ref{fig:mercedes} (a) and (b),
where the black circle in panel (a) shows a
gate-voltage vector with half-integer components, 
whereas the black solid loop in panel (b)
shows the corresponding ground-state degeneracy 
pattern, i.e., a nodal loop. 
The black circle in panel (a) is a transition point, 
and the corresponding nodal loop 
is an extended degeneracy pattern,
analogous to the sphere in the magnetic field space
seen in Fig.~\ref{fig:doubledot}c.
The circuit parameters used to obtain this result
are 
$(
E_{J,01},
E_{J,02},
E_{J,03},
E_{J,12},
E_{J,13},
E_{J,23}
)/h = (2,4,6,3,3,6)\,\text{GHz}$,
and
$(
C_{01},
C_{02},
C_{03},
C_{12},
C_{13},
C_{23}
)/h = (2,1,2,3,4,3)\,\text{fF}$.
Furthermore, capacitances between terminals 1, 2, 3 and their respective control gates are 
$C_{\text{g},1} = C_{\text{g},2} = C_{\text{g},3} = 0.1\, \text{fF}$.
(Numerical techniques used to obtain this result can be found in the Appendix.)

The nodal loop serves as a wormhole for the Weyl points, 
enabling their teleportation.
This is illustrated in Fig.~\ref{fig:mercedes} (a) and (b).
In (a), a continuous trajectory 
$\blacksquare \to \circ \to +$
including the transition 
point ($\circ$) is shown (green line). 
Here, the incoming and outgoing
paths are straight, enclosing a finite angle.
Formally, the trajectory in control space is defined as 
$\vec{n}_\text{g}(t) = \vec n_{\text{g}0}-t \vec{e}_\blacksquare$
for $t<0$, 
$\vec{n}_\text{g}(t) = \vec n_{\text{g}0}+t \vec{e}_+$
for $t\geq 0$, with
$\vec e = \vec e_{\blacksquare} = (3,8,6)/\sqrt{109}$,
$\vec e = \vec e_{+} = (3,8,-6)/\sqrt{109}$, 
and
$t\in [-0.1,0.1]$.
The corresponding motion of 4 ground-state 
Weyl points in the configuration space is shown in
(b), where the red (blue) 
color denotes the $+1$ ($-1$) topological charge
(ground-state Chern number) of the Weyl point.
Even though the path in the control space is continuous (Fig.~\ref{fig:mercedes}a),
the Weyl-point positions in the configuration space
suffer a sudden jump as the control-space trajectory
crosses the transition point (Fig.~\ref{fig:mercedes}b).

In the two-spin model described above, we have demonstrated the $1/E$ divergence of motional susceptibility of the Weyl points. 
We now identify the same type of divergence in  this Weyl Josephson circuit model, using
the results in Fig.~\ref{fig:mercedes}c.
For each of the two directions of Fig.~\ref{fig:mercedes}a,
i.e., the ones denoted by the green square and the green cross, we numerically compute the greatest singular value of the motional susceptibility matrix of a selected Weyl point as the control vector is changed according to $\vec n_\text{g}(t) = \vec n_{\text{g}0} + t \vec e$ with $t \in [10^{-4},0.01]$, for two cases: $\vec{e} = \vec{e}_\blacksquare$ and $\vec{e} = \vec{e}_+$.
Figure \ref{fig:mercedes}c shows the obtained data points, along with the two single-parameter fits to the data (solid lines). 
These results reveal that in the vicinity of the transition point (that is, for $t\to 0$), all of these singular values show a $1/t$ divergence. 
We have checked that this behavior holds for all Weyl points created from the nodal loop, for various Hamiltonian parameter values and directions in control space.

Based on our results for the two-spin model and the Weyl Josephson circuit model, we conjecture that this $1/t$ divergence of Weyl-point motional susceptibility is a generic feature in the vicinity of a nodal loop and a nodal surface, irrespective of the specific physical setup. 
We pose it as an open problem to rigorously identify the preconditions of this scaling law.

In fact, the divergent motional susceptibility of the Weyl points in the vicinity of the nodal surface or nodal line can be regarded as a side effect of teleportation. 
A simple argument is as follows. 
Consider a circular path connecting the control points
$\vec{n}_{\text{g}0} + t \vec e_\blacksquare$
and
$\vec{n}_{\text{g}0} + t \vec e_+$, centered
around the transition point $\vec{n}_{\text{g}0}$. 
The path length approaches zero as $t\to 0$, being proportional to $t$.
We compare this with the corresponding Weyl-point path length in the configuration space, by inspecting, e.g., the top left Weyl point in Fig.~\ref{fig:mercedes}b. 
The Weyl-point path length does not converge to zero as $t\to 0$; instead, it converges to the path length between the meeting points of the Weyl-point trajectory and the nodal loop (i.e., the \emph{teleportation length}), which is finite. 
This simple consideration explains that in the vicinity of the transition point, a small generic change in the control vector implies a large jump of the Weyl points; furthermore, it provides an interpretation of the $1/t$ type divergence of the motional susceptibility. 

\emph{Conclusions.} In conclusion, we have argued that nodal loops and nodal surfaces of parameter-dependent quantum systems serve as wormholes for Weyl points, allowing their teleportation. 
Furthermore, as the control parameters approach the transition point where the degeneracy points form a loop or a surface, the pace of motion of the Weyl points in the configuration space diverges, following a simple scaling law.
A special case of the teleportation effect was found in superfluid $^3$He-A~\cite{nissinen2018} where two oppositely charged Weyl points switch places through a nodal loop.
We exemplified more general teleportation patterns on a two-spin model and a Weyl Josephson circuit, and we expect that they generalize naturally to many other settings, including electronic \cite{Armitage,ChenFangChinese2016,YingMingXie,Wilde,li2021}, photonic \cite{WenlongGao,BiaoYangPhotonicWeyl}, phononic \cite{FengLiPhononicWeyl}, magnonic \cite{Romhanyi}, and synthetic \cite{Riwar} band structures.

\emph{Author contributions.}
Gy.~F. performed analytical and numerical calculations, and produced the figures.
Gy.~F. and A.~P. wrote the initial draft of the manuscript.
A.~P. acquired funding, and managed the project.
G.~P. consulted on differential-geometric aspects of the work.
All authors contributed to the formulation of the project, discussed the results and took part in writing the manuscript.

\acknowledgments
We acknowledge fruitful discussions and correspondence with J. Asb\'oth, L. Bretheau, V. Fatemi, Y. Qian, A. Schnyder, and H. Weng. 
This research was supported by the Ministry of Innovation and Technology (MIT) and the National Research, Development and Innovation Office (NKFIH) within the Quantum Information National Laboratory of Hungary and the Quantum Technology National Excellence Program (Project No. 2017-1.2.1-NKP-2017-00001), by the 
NKFIH
fund TKP2020 IES (Grant No. BME-IE-NAT) under the auspices of the MIT, and by the
NKFIH through the OTKA Grants FK 124723 and FK 132146.
D.~V. was supported by the Swedish Research Council (VR) and the Knut and Alice Wallenberg Foundation.

\bibliography{references.bib}

\newpage
\appendix

\clearpage
\section{Exchange rotation in the interacting two-spin model}

Here, we provide a heuristic estimate for the exchange
rotation angle $\gamma E$ included in 
the interacting two-spin model of the main text. 

The exchange interaction in Eq.~\eqref{eq:exchangerotation}
contains a rotation matrix of angle $\gamma E$.
To estimate $\gamma$ in a realistic setting, 
we consider electrons in an InAs nanowire
($m_\text{eff}=0.023\, m_\text{e}$), subject to 
an electric field that is assumed to be homogeneous. 
We identify the exchange rotation angle with the
spin rotation angle via
\begin{equation}
\gamma E = 2\pi \ell_\text{d}/\ell_\text{so}    
\end{equation}
of a free conduction electron in the nanowire
as it traverses the interdot distance 
$\ell_\text{d}$
of the double quantum
dot system. 
The spin-orbit length $\ell_\text{so}$, according to
\cite{ScherublRashba}, can be expressed as
\begin{equation}
    \ell_\text{so} = 
    \frac{\hbar^2}{m_\text{eff} e \alpha_0 E},
\end{equation}
with $\alpha_0 = 1.17 \, \text{nm}^2$.

The above relations imply
\begin{equation}
\gamma = \frac{2\pi l_\text{d} m_\text{eff} e \alpha_0}{\hbar^2}
\approx 222 \, \text{nm}/\text{V},
\end{equation}
where an interdot distance of 
$\ell_\text{d} = 100\, \text{nm}$ was used.
That is, an electric field $E = 1 \, \mu\text{m}/\text{V}$ 
implies an exchange rotation of $\gamma E = 0.222$ radians, 
approximately 13 degrees.

\section{The capacitance matrix of the Weyl Josephson circuit}

The capacitance matrix \cite{vanderWielRMP} of the Weyl Josephson circuit described in the main text reads:
\bean
C=\begin{pmatrix}
C_1&-C_{12}&-C_{13}\\
-C_{12}&C_2&-C_{23}\\
-C_{13}&-C_{23}&C_3
\end{pmatrix},
\eean
where
\begin{subequations}
\bean
C_1 &=& C_{01} + C_{12} + C_{13} + C_{\text{g1}},\\
C_2 &=& C_{02} + C_{12} + C_{23} + C_{\text{g2}},\\
C_3 &=& C_{03} + C_{13} + C_{23} + C_{\text{g3}}.
\eean
\end{subequations}

\section{Effective $\mathcal{PT}$ symmetry protecting the nodal loops of the Weyl Josephson circuit}
\label{app:ptsymmetry}

The Josephson-circuit Hamiltonian $\hat{H}(\vec n_g,\vec \varphi)$ has time-reversal symmetry in the absence of magnetic fluxes:
\bean
\label{eq:timereversal}
\mathcal{T} \hat{H}(\vec n_\text{g}, 0) \mathcal{T}^{-1} = \hat{H}(\vec n_\text{g}, 0),
\eean
where $\mathcal{T} = \mathcal{K}$ is the complex conjugation in the charge basis \cite{Fatemi}, and hence $\mathcal{T}^2=1$.
Furthermore, for nonzero magnetic flux, the relation \eqref{eq:timereversal} generalizes as
\bean
\mathcal{T} \hat{H}(\vec n_\text{g},\vec \varphi) \mathcal{T}^{-1} = \hat{H}(\vec n_\text{g}, -\vec \varphi).
\eean
In the charge basis, this is translated to 
\bean
H^{\ast}_{\vec n,\vec n'}(\vec n_\text{g},\vec \varphi) = H_{\vec n,\vec n'}(\vec n_\text{g},-\vec \varphi).
\eean

Other symmetries of $\hat{H}$ are charge inversion
\bean
H(\vec n_\text{g}, \vec \varphi)_{\vec n, \vec n'} = H(-\vec n_\text{g}, -\vec \varphi)_{-\vec n, -\vec n'},
\eean
and discrete charge translation symmetry
\bean
H(\vec n_\text{g}, \vec \varphi)_{\vec n, \vec n'} = H(\vec n_\text{g} + \vec m,\vec \varphi)_{\vec n + \vec m, \vec n' + \vec m}
\eean
where $\vec m$ is an arbitrary integer offset charge vector.

The charge inversion and charge translation together gives 
\bean
H(\vec n_\text{g}, \vec \varphi)_{\vec n, \vec n'} = H(\vec m-\vec n_\text{g}, -\vec \varphi)_{\vec m-\vec n,\vec m - \vec n'}.
\eean
This relation implies that for fixed  integer or a half-integer offset charge vector $\vec n_g=\frac{\vec m}{2}$, it holds that
\bean
\label{eq:combined}
H\left(\frac{\vec m}{2},\vec \varphi\right)_{\vec n, \vec n'} =H\left(\frac{\vec m}{2}, -\vec \varphi\right)_{\vec m-\vec n,\vec m - \vec n'}.
\eean
This relation can be expressed by the charge inversion operator $\mathcal{P}^{(\vec m /2)}$, whose matrix elements in the charge basis are
\bean
P^{(\vec m/2)}_{\vec n, \vec n'} = \delta_{\vec n, \vec m - \vec n'}.
\eean
With this definition, we rewrite Eq.~\eqref{eq:combined} as
\bean
\mathcal{P}^{(\vec m/2)} \hat{H}\left(\frac{\vec m} 2, \vec \varphi \right) \left(\mathcal{P}^{(\vec m/2)}\right)^{-1} = \hat{H}\left(\frac{\vec m} 2, - \vec \varphi \right).
\eean
inverting the charge with respect to $\frac{\vec m}{2}$.
This is an exact symmetry even when the charge basis is restricted to a finite interval symmetrically around $\frac{\vec m}{2}$.

Henceforth, we fix the charge inversion point as $\frac{\vec m} 2$, and suppress it in the formulas below.
Then, the combination of charge inversion and time-reversal symmetry results in
\bean
\label{eq:secondtimereversal}
(\mathcal{PT}) \hat{H}\left(\vec \varphi\right) \mathcal{(PT)}^{-1} = \hat{H}\left(\vec \varphi\right),
\eean
i.e., we have identifyed an antiunitary symmetry $\mathcal{PT}$ with $(\mathcal{PT})^2 = 1$ that restricts the Hamiltonian at every value of $\vec \varphi$ for a fixed $\vec n_g=\frac{\vec m}{2}$ integer or a half-integer offset charge vector.
Carrying out a basis transformation from the charge basis by the unitary $U = \sqrt{\mathcal{P}}$, $\mathcal{PT}$ is represented as complex conjugation, making the transformed $H(\vec \varphi)$ real.
The co-dimension for real valued symmetric matrices to be two-fold degenerate is 2 (see, e.g., Ref.~\cite{Neumann}), meaning that the level crossings generally appear as 1 dimensional space curves in the 3 dimensional $\vec \varphi$ parameter space.

To show a specific example, we consider the truncated $8\times 8$ Hamiltonian of the Josephson circuit in the charge basis
$\{\ket{000},\ket{100},\ket{010},\ket{001},\ket{111},\ket{011},\ket{101},\ket{110}\}$, which reads
\begin{flalign}
\label{eq:H8x8}
&H_{8\times 8}(\vec n_g, \vec \varphi)=&\nonumber\\
&=\begin{pmatrix}
\epsilon_{000}&h_{01}&h_{02}&h_{03}&0&0&0&0\\
h_{01}&\epsilon_{100}&h_{12}^\ast &h_{31}&0&0&h_{03}&h_{02}\\
h_{02}&h_{12}&\epsilon_{010}&h_{23}^\ast &0&h_{03}&0&h_{01}\\
h_{03}&h_{31}^\ast &h_{23}&\epsilon_{001}&0&h_{02}&h_{01}&0\\
0&0&0&0&\epsilon_{111}&h_{01}&h_{02}&h_{03}\\
0&0&h_{03}&h_{02}&h_{01}&\epsilon_{011}&h_{12}&h_{31}^\ast \\
0&h_{03}&0&h_{01}&h_{02}&h_{12}^\ast &\epsilon_{101}&h_{23}\\
0&h_{02}&h_{01}&0&h_{03}&h_{31}&h_{23}^\ast &\epsilon_{110}
\end{pmatrix},&
\end{flalign}
where
\bean\label{eq:H8x8elements}
\epsilon_{abc}&=&E_\text{C}[(a,b,c)^{\text T}-\vec n_\text{g}]\cdot c^{-1} [(a,b,c)^{\text T}-\vec n_\text{g}]\\
h_{\alpha \beta}&=&-\frac{1}{2}E_{\text{J},\alpha \beta}e^{-i\gamma_{\alpha \beta}}.
\eean
At the time-reversal offset charge point \mbox{$\vec n_{\text{g}}=\left(\frac{1}{2},\frac{1}{2},\frac{1}{2}\right)^{\text T}\equiv \vec{\frac{1}{2}}$}, the diagonal elements have the property
\bean\label{eq:chargesymm}
\epsilon_{a,b,c}=\epsilon_{1-a, 1-b, 1-c},
\eean
yielding the $\mathcal{PT}$ symmetry
\bean\label{eq:H8x8symm}
PH^{\ast}_{8\times 8}\left(\vec \varphi\right)P^{-1}=H_{8\times 8}\left(\vec \varphi\right),
\eean
with $P=\sigma_x\otimes\mathds{1}_{4\times 4}$. The corresponding unitary matrix  \mbox{$U=\sqrt{i/2}\left(\mathds{1}_{2\times 2}-i\sigma_x\right)\otimes\mathds{1}_{4\times 4}$} transforms the Hamiltonian matrix to be real-valued
\bean\label{eq:H8x8real}
UH_{8\times 8}\left(\vec \varphi\right)U^{-1}=\left[UH_{8\times 8}\left(\vec \varphi\right)U^{-1}\right]^{\ast}.
\eean

\section{Weyl Josephson circuit: Numerical techniques}

The calculations of the Weyl Josephson circuits were done numerically, with the aid of analytical techniques. The starting point of the calculation was the truncated matrix representation of the Hamiltonian of Eq.~(10) of the main text. The matrix representation was obtained by projecting the Hamiltonian onto the subspace spanned by the charge basis states $\ket{n_1,n_2,n_3}$, with \mbox{$n_1, n_2, n_3 \in\{-1,0,1,2\}$}, yielding a $4^3\times 4^3 = 64 \times 64$ matrix.

This charge interval is chosen to be symmetric to $1/2$, to respect the symmetry of Eq.~\eqref{eq:secondtimereversal} protecting the nodal loop at \mbox{$\vec n_{\text{g}0}=(0.5,0.5,0.5)$}. The range of the charge states was chosen to be large enough to describe the physical system accurately, and small enough to save computational time.

\begin{figure}
	\begin{center}
		\includegraphics[width=0.9\columnwidth]{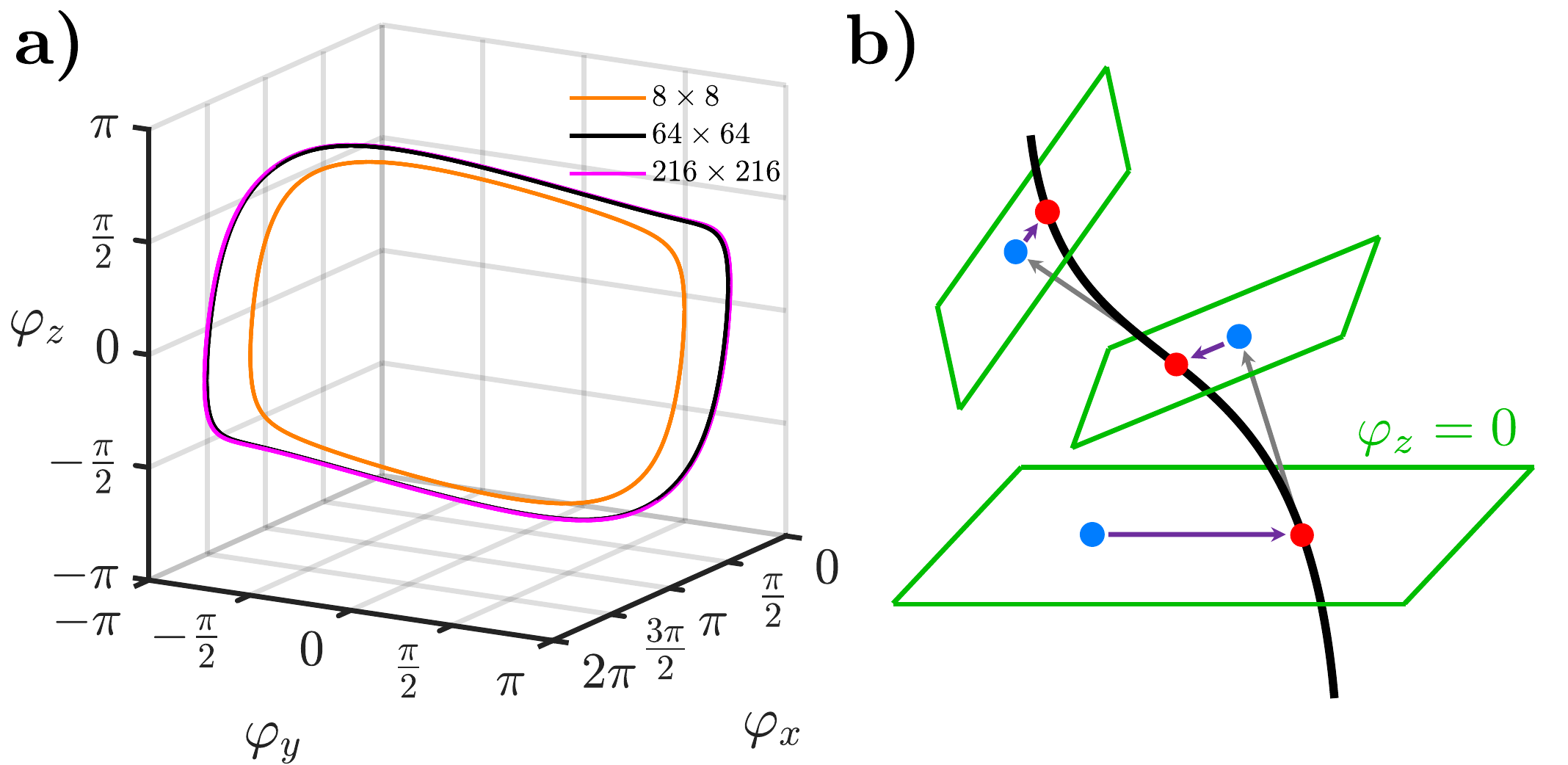}
	\end{center}
	\caption{{\bf Nodal loop in the Weyl Josephson junction.} a) Numerical identification of nodal loops and Weyl points was carried out by truncating the infinite-dimensional Hilbert space of the Weyl Josephson circuit. Larger Hamiltonian matrices give more accurate results: the corresponding nodal loops converge. b) Numerical search for the points (red) of the nodal loop (black). The search consists of two nested iterations: the inner iteration (purple) searches for the intersection of the nodal loop with a plane (green). After the intersection is found, the outer iteration makes a step of a finite distance in the tangential direction of the nodal loop (grey) and determines the new searching plane as the perpendicular plane passing through the new point (blue). The first inner iteration searches in the $\varphi_z=0$ plane.
	\label{fig:loopapp}}
\end{figure}

To check the error due to the truncation of the Hamiltonian, we have compared the numerically identified nodal loops for smaller and larger matrix dimensions, see Fig.~\ref{fig:loopapp}a. 
Besides the $64 \times 64$ truncation size, the figure shows results with the \mbox{$8\times 8$} and \mbox{$216\times 216$} Hamiltonians, corresponding to the charge ranges \mbox{$0\leq n_{1,2,3}\leq 1$} and \mbox{$-2\leq n_{1,2,3}\leq 3$}, respectively. We found that all truncated Hamiltonians have the same qualitative behaviour, i.e., presence of nodal loop, Weyl point teleportation, and susceptibility divergence. Moreover, the quantitative difference between the results from the two larger Hamiltonians is negligible, e.g., the distance between the points of the nodal loops in the $\varphi_z=0$ plane is 0.036 rad (see Fig.~\ref{fig:loopapp}a).

In the following subsections, we present the numerical techniques for characterizing the nodal loop and searching for the Weyl points. We use a generic description where $\vec p$ is the 3-component configurational parameter in the configurational space where the Weyl points and the nodal loop are present, and $\vec q$ is the $k$-component control parameter which can tune system. 
The correspondence between this notation and the notation of gate voltages and magnetic fluxes In the Weyl Josephson circuit reads:
\bean\label{eq:pq_parameters}
\vec p&\equiv& \vec \varphi=(\varphi_x,\varphi_y,\varphi_z),\\
\vec q&\equiv&\vec n_{\text g}=(n_{\text g,1},n_{\text g,2},n_{\text g,3}).
\eean

\subsection{Effective $2\times 2$ Hamiltonian near two-fold degeneracies}
Searching for the degeneracies of a large Hamiltonian can be computationally demanding. To ease the problem, a useful technique is to reduce the Hamiltonian to a low-dimensional effective model with Schrieffer--Wolff transformation. Consider the Hamiltonian $H(\vec p,\vec q)$ depending on the 3 dimensional configurational space $\vec p$ and the $k$ dimensional $\vec q$ control space. At $(\vec p_0$,$\vec q_0)$, the $i$th and $(i+1)$th energies, to be denoted as $E_i(\vec p_0,\vec q_0)$ and $E_{i+1}(\vec p_0,\vec q_0)$, are close together and well separated from the other energy levels.

In our method to find a degeneracy point in the vicinity of such a quasi-degenerate point ($\vec p_0, \vec q_0$), a useful ingredient is to expand the Hamiltonian in linear order as
\bean\label{eq:hlinear}
&H&^{(1)}(\vec p_0+\delta \vec p,\vec q_0+\delta \vec q)=\nonumber\\
&H&(\vec p_0,\vec q_0)+\left.\frac{\partial H}{\partial \vec{p}} 
    \right|_{\vec p_0,\vec q_0}\cdot \delta \vec p+\left.\frac{\partial H}{\partial \vec{q}} 
    \right|_{\vec p_0,\vec q_0}\cdot \delta \vec q,
\eean
and project it to the subspace of the quasi-degenerate levels with $P_0=\ket{\Psi_i}\bra{\Psi_i}+\ket{\Psi_{i+1}}\bra{\Psi_{i+1}}$, where $\ket{\Psi_i}$ and $\ket{\Psi_{i+1}}$ are the $i$th and $(i+1)$th energy eigenstates of $H(\vec p_0,\vec q_0)$. 
This projection yields the following $2\times 2$ effective Hamiltonian:
\bean\label{eq:hefflinear}
H^{(1)}_\text{eff}&=&P_0 H^{(1)} P_0\nonumber\\
&\equiv&\vec\sigma\cdot\vec\Delta^{(1)} = \vec \sigma\cdot\left[\vec \Delta_0 + g_{\vec p}\delta\vec p + g_{\vec q}\delta\vec q\right].
\eean
Here, $\vec \sigma = (\sigma_x,\sigma_y,\sigma_z)$ is the vector of Pauli operators acting on the quasi-degenerate subspace. 
We neglected the $\sigma_0$ term in the expansion, since because it does not affect the splitting between the two states. 

By the above definition, the only nonzero component of the \emph{polarization vector} $\vec \Delta_0$ is its third component. 
Furthermore, we call $g_{\vec p}$ and $g_{\vec q}$ as the configurational-space and control-space \emph{effective $g$-tensors}, which are a real valued $3\times 3$ and $3\times k$ matrices:
\bean\label{eq:heffcoeff}
    \left[\Delta_0(\vec p_0,\vec q_0)\right]_3&=&\frac{1}{2}\left[
    E_i(\vec p_0,\vec q_0)-E_{i+1}(\vec p_0,\vec q_0)\right],\nonumber\\
    \left[g_{\vec p}(\vec p_0,\vec q_0)\right]_{\alpha,\beta}
    &=&
    \frac{1}{2}\text{Tr}\left(
    \sigma_\alpha 
    P_0 
    \left. 
        \frac{\partial H}{\partial p_\beta} 
    \right|_{\vec p_0,\vec q_0}
    P_0
    \right),\\
    \left[g_{\vec q}(\vec p_0,\vec q_0)\right]_{\alpha,\beta}
    &=&
    \frac{1}{2}\text{Tr}\left(
    \sigma_\alpha 
    P_0 
    \left. 
        \frac{\partial H}{\partial q_\beta} 
    \right|_{\vec p_0,\vec q_0}
    P_0
    \right).\nonumber
\eean
Note that the polarization vector and the effective $g$-tensors are basis dependent. The SU(2) transformation of the basis multiplies them with a corresponding SO(3) matrix from the left. If the transformation is only the change of the phase difference between the states, the corresponding orthogonal matrix $R_z(\varphi)$ describes a rotation around the $z$ axis. Later, we use these basis dependent quantities to calculate the location and the susceptibility matrix of the Weyl points which are independent from the basis choice. This independence can be inferred from the formulas.

When searching for a degeneracy point $\vec p_0 + \delta \vec p$ in the vicinity of $p_0$, for a fixed value of $\vec q_0 + \delta \vec q$, an approximate result can be obtained by requiring that the energy gap of the effective Hamiltonian of Eq.~\eqref{eq:hefflinear} should vanish:
\bean\label{eq:DeltaZero}
\vec \Delta^{(1)}=\vec 0.
\eean
The solution of this equation for $\delta \vec p$ provides the approximate position $\vec p_0 + \delta \vec p$ of the degeneracy point.

It is worth noting that the second order Schrieffer--Wolff transformation is more than the projection of a second order expansion of the Hamiltonian to its near-degenerate subspace.

\subsection{Finding Weyl points}
The Weyl points are the generic point-like degeneracies in the three dimensional configurational space (containing the points $\vec p$), for a fixed control parameter vector $\vec q_0$. 
We neglect $\vec q_0$ in the arguments for brevity. 
The equation providing an approximate position of the degeneracy is that the effective Hamiltonian should vanish, i.e., 
\bean\label{eq:Heff}
H^{(1)}_\text{eff}\left(\vec p_0+\delta\vec p\right)=\vec \sigma\cdot\left[\vec \Delta_0\left(\vec p_0\right)+g_{\vec p}\left(\vec p_0\right)\delta \vec p\right]=0.
\eean
Rearranging yields
\bean\label{eq:linear}
g_{\vec p}\left(\vec p_0\right)\delta \vec p=-\vec \Delta_0\left(\vec p_0\right),
\eean
which is solved using the inverse of the $g$-tensor
\bean\label{eq:weylsolution}
\delta \vec p=-g_{\vec p}^{-1}\left(\vec p_0\right)\vec \Delta_0\left(\vec p_0\right).
\eean

The resulting $\vec p_0+\delta \vec p$ is not an exact Weyl point in general, but it is closer to the Weyl point than $\vec p_0$. Following this scheme, we iterate the process with the following formula
\bean\label{eq:weyliteration2}
\vec p_{n+1}=\vec p_{n}-g_{\vec p}\left(\vec p_n\right)^{-1}\vec \Delta_0(\vec p_n),
\eean
until the energy splitting is sufficiently small; in our numerics, we used the threshold of $10^{-12}$ GHz. 
The starting point $\vec p_0$ needs to be close to the exact Weyl point. 
When searching for an ordinary Weyl point with the above iteration, the effective $g$-tensor is non-singular throughout the iteration, as it is non-singular in the Weyl point. 
Hence the inverse used in the iteration of Eq.~\eqref{eq:weyliteration2} does exist.

\subsection{Finding the nodal loop}
In the presence of a nodal loop, our goal is to numerically locate a discrete set of its points that allows us to draw the loop.
To achieve that, we intersect the latter with planes in the configurational space, as shown in Fig.~\ref{fig:loopapp}b. 
From the initial point $\vec p$ (see, e.g., the blue point in the $\varphi_z = 0$ plane in Fig.~\ref{fig:loopapp}b),  we restrict the displacement $\delta \vec p$ as
\bean\label{eq:nodalplane}
\delta \vec p=c_1 \vec v_1+c_2 \vec v_2=
\underbrace{
\left(
	\begin{array}{c|c}
		\vec v_1 & \vec v_2
	\end{array}
\right)}_V
\underbrace{
\begin{pmatrix}
c_1\\
c_2
\end{pmatrix}}_{\vec c},
\eean
where $\vec v_1$ and $\vec v_2$ are two arbitrary vectors in the 3D configurational space that span the plane. The condition for the Weyl point in the linearized effective Hamiltonian in Eq.~\eqref{eq:linear} changes to
\bean\label{eq:linear2}
\underbrace{g_{\vec p}\left(\vec p_0\right)V}_{g_V}\vec c &=& -\vec \Delta_0(\vec p_0).
\eean
At first sight this is an overdetermined linear set for $c_1$ and $c_2$, since it contains 3 conditions but only 2 variables. However, because of the $\mathcal{PT}$ symmetry discussed in App.~\ref{app:ptsymmetry}, there is a basis where the effective Hamiltonian is real-valued. In this basis, the second component of Eq.~\eqref{eq:linear2} simplifies to the identity $0=0$, meaning that there is a unique solution for the equation.

One way to solve Eq.~\eqref{eq:linear2} is to look for the vector $\vec c$ where $|g_V\vec c + \vec \Delta_0|^2$ is minimal. 
Note that at the minimum point, the minimum value is actually zero. 
Equating the derivative of 
$|g_V\vec c + \vec \Delta_0|^2$
with respect to $\vec c$ to zero yields 
\bean\label{eq:linear3}
g_V^{\text T}g_V\vec c = -g_V^{\text T}\vec \Delta_0.
\eean
Note that Eq.~\eqref{eq:linear3} can also be obtained by multiplying Eq.~\eqref{eq:linear2} with $g_V^{\text T}$ from the left.
Now we can solve Eq.~\eqref{eq:linear3} using the inverse of the $2\times 2$ matrix $g_V^{\text T} g_V$ as
\bean\label{eq:nodalsolution}
\vec c &=& -\left(g_V^{\text T}g_V\right)^{-1}g_V^{\text T}\vec \Delta_0.
\eean

Following the spirit of the iteration at the end of the previous section, the analogous iteration to find the intersection of the nodal loop and the considered plane is defined as
\bean\label{eq:nodaliteration}
&\vec p&_{n+1} = \vec p_n + \delta \vec p 
= \vec p_n + V \vec c = \\
&\vec p&_{n}-V\left[V^{\text T}g^{\text T}_\vec p\left(\vec p_n\right)g_\vec p\left(\vec p_n\right) V\right]^{-1}V^{\text T}g^{\text T}_{\vec p}\left(\vec p_n\right)\vec \Delta_0\left(\vec p_n\right).\nonumber
\eean
where we have used Eqs.~\eqref{eq:nodalplane}, \eqref{eq:linear2}, and \eqref{eq:nodalsolution}.

The first point of the loop is searched in the $\varphi_z=0$ plane. Then every starting point searching the points of the nodal loop is \mbox{0.01} rad distance in the tangential direction of the nodal loop from the previous point. The search is restricted to the plane perpendicular to the step (Fig.~\ref{fig:loopapp}). The tangent vector $\vec v$ of the nodal loop is determined by the $g$-tensor as $g_{\vec p}\vec v=0$, which means that the splitting is zero in the linear order in that direction.

\subsection{Susceptibility matrix of ordinary Weyl points}

Here, we express the motional susceptibility matrix $\chi$ introduced in Eq.~(4) of the main text with the configurational- and control-space effective $g$-tensors $g_{\vec p}$ and $g_{\vec{q}}$, respectively. 
The susceptibility matrix $\chi$ describes the motion of the Weyl points in the configurational parameter space $\vec p$ upon changing the control parameters $\vec q$. The linear effective Hamiltonian expanded at a Weyl point does not have a constant term:
\bean
\vec\Delta^{(1)}(\vec p,\vec q)=g_{\vec p}(\vec p_0,\vec q_0)\delta\vec p+g_{\vec q}(\vec p_0,\vec q_0)\delta\vec q.
\eean
The condition for Weyl points reads $\vec\Delta^{(1)}=0$. Rearranging this equation and acting with $g_{\vec p}^{-1}$ from the left gives
\bean
\delta\vec p=-\left(g_{\vec p}^{-1}g_{\vec q}\right)\delta\vec q=\chi(\vec p_0,\vec q_0)\delta\vec q,
\eean
which determines the linear order displacement $\delta \vec p$ of the Weyl point for the arbitrary perturbation $\delta \vec q$ in the control space, which is in fact the definition of the susceptibility matrix.

Notice that the susceptibility matrix is independent from the choice of basis in the two-fold degenerate eigenspace at the Weyl point. Changing the basis multiplies the $g$-tensors $g_{\vec p}$ and $g_{\vec q}$ with the same orthogonal matrix from the left, and this exactly cancels out in the expression $g_{\vec p}^{-1}g_{\vec q}$. 

\section{Survivor Weyl points close to the transition point}
\begin{figure}
	\begin{center}
		\includegraphics[width=0.9\columnwidth]{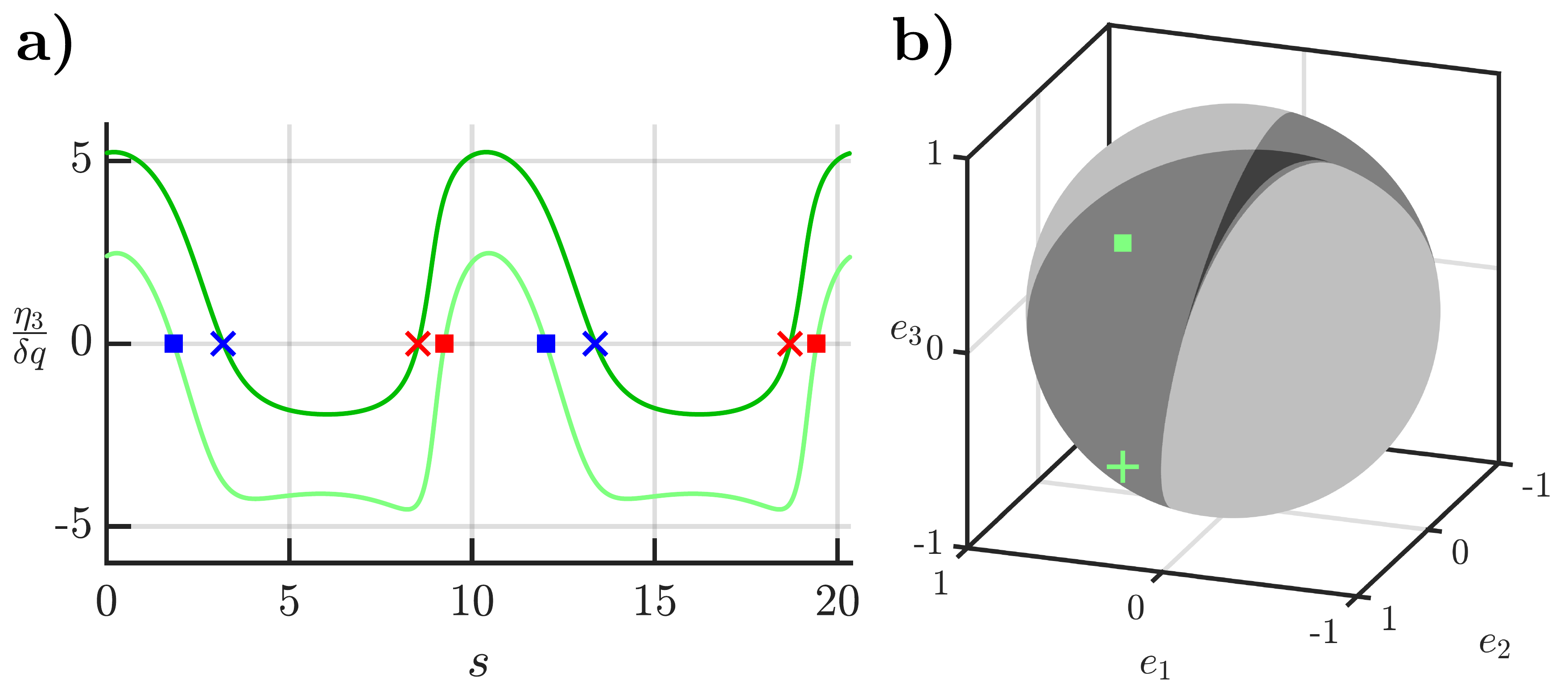}
	\end{center}
	\caption{{\bf Survivor Weyl points of the broken nodal loop.} a) The zeros of the function $\eta_3(s)$ (third component of right hand side of Eq.~\eqref{eq:sequation}) determines where the survivor Weyl points are located for an infinitesimal perturbation. For perturbations in different directions of the control space, the function $\eta_3(s)$ is different, causing the Weyl-point teleportation. b) The direction of the perturbation also determines the number of the survivor Weyl points. The pale gray region has no Weyl points, the medium gray has 4 Weyl points and the dark grey region has 8 Weyl points. The perturbations used in the main text is in the 4-point region. 
	For both markers (square, cross) in b), the arc length parameters of the survivor Weyl points are depicted by the same markers on a). 
	\label{fig:survivorweyl}}
\end{figure}

In this section, we derive the location of the Weyl points that `survive' upon an infinitesimal perturbation of a nodal loop.
Moreover, we show that the number of survivor Weyl points depends on \emph{how} the nodal loop is perturbed: for the example considered, the number of survivor Weyl points is either 0, 4, or 8. 

The nodal loop $\vec p_0(s)$ in the configurational parameter space can be parametrized by its arc length $s$ and its vicinity can be parametrized by the $\delta\vec p_\perp$ perpendicular displacement from the nodal loop. The perturbed effective Hamiltonian reads
\bean
\vec \Delta^{(1)}=g_{\vec p}(s)\delta\vec p_\perp+g_{\vec q}(s)\delta\vec q,
\eean
where we omitted the $\vec q_0$-dependence in the arguments of the $g$-tensors for simplicity because the nodal loop corresponds to only specific discrete $\vec q_0$ values. The survivor Weyl points are located at the arc length parameter $s$ where the condition 
\begin{equation}
\label{eq:survivor}
    \vec \Delta^{(1)}=0
\end{equation} has solution for $\delta \vec p_\perp$. The $g_{\vec p}(s)$ configurational-space effective $g$-tensor is a singular matrix for the points of the nodal loops, hence we can not solve Eq.~\eqref{eq:survivor} using the inverse of it. Instead, we use the singular value decomposition of $g_{\vec p}(s)$ to rewrite Eq.~\eqref{eq:survivor} as
\bean
U(s)\Sigma(s)V(s)^{-1}\delta\vec p_\perp&=&-g_{\vec q}(s)\delta\vec q\\
\Sigma(s)\underbrace{V(s)^{-1}\delta\vec p_\perp}_{\delta \tilde {\vec p}_\perp}&=&-\underbrace{U(s)^{-1}g_{\vec q}(s)\delta\vec q}_{\vec \eta(s,\delta \vec q)}
\label{eq:sequation}
\eean
where $\Sigma = \text{diag}(\Sigma_1,\Sigma_2,0)$ is a diagonal matrix and $U$ and $V$ are $3 \times 3$ orthogonal matrices.
The third column of $\Sigma$ is zero, therefore the third component of the left hand side is also zero. Therefore, the third component of Eq.~\eqref{eq:sequation}, that is, 
\bean
\eta_3(s_0,\delta \vec q)=0,
\eean
can be used to find the arc length parameters $s_0$ where Weyl points appear upon applying the perturbation $\delta \vec q$.
It is worth noting that $\vec \eta (s,\delta \vec q)$ is independent of the  choice of the basis of the two-dimensional degenerate subspace at the Weyl point. 
The first-order perpendicular-to-nodal-loop of the survivor Weyl points is given as
\bean
\delta\vec p_\perp=-g_\vec p^{+}\left(s_0(\delta \vec q)\right)g_\vec q\left(s_0(\delta \vec q)\right)\delta \vec q,
\eean
where $g_\vec p^{+}=V\Sigma^{+} U^{-1}$ is the Moore--Penrose inverse of the $g$-tensor with $\Sigma^{+} = \text{diag}(\Sigma_1^{-1},\Sigma_2^{-1},0)$. There is also a first-order parallel-to-nodal-loop motion of the survivor Weyl points, but determining that requires a second order calculation.

This derivation indicates the Weyl-point teleportation: for infinitesimal perturbations $\delta\vec q$ in different directions the function $\eta_3(s,\delta\vec q)$ has roots at different arc length values (see Fig.~\ref{fig:survivorweyl}a), meaning that different points of the nodal loop survive. For infinitesimally small perturbations $\delta \vec q=\vec n\delta q$ only the direction $\vec n$ describes the location of the surviving Weyl points. Their number also depends on the direction of the perturbation which gives a phase diagram on the unit sphere (see Fig.~\ref{fig:survivorweyl}b).

\end{document}